\documentclass[]{spie}  %>>> use for US letter paper

\usepackage[]{graphicx}
\usepackage{amssymb}
\usepackage{amsmath}
\usepackage{graphicx}
\usepackage{bm}
\usepackage{amsfonts}

\def\be{\begin{equation}} 
\def\ee{\end{equation}}

\title{Characterizing a Four-Qubit Planar Lattice for Arbitrary Error Detection} 

\author{Jerry M. Chow\supit{a}, Srikanth J. Srinivasan\supit{a}, Easwar Magesan\supit{a}, A. D. C\'orcoles\supit{a}, David W. Abraham\supit{a}, Jay M. Gambetta\supit{a}, Matthias Steffen\supit{a}
\skiplinehalf
\supit{a}IBM T.J. Watson Research Center, Yorktown Heights, NY, U.S.A;
}

\authorinfo{Further author information: (Send correspondence to J.M.C.)\\J.M.C.: E-mail: chowmj@us.ibm.com, Telephone: 1 914 945 2695}

\begin{document} 
\maketitle 

\begin{abstract}
Quantum error correction will be a necessary component towards realizing scalable quantum computers with physical qubits. Theoretically, it is possible to perform arbitrarily long computations if the error rate is below a threshold value. The two-dimensional surface code permits relatively high fault-tolerant thresholds at the $\sim$1\% level, and only requires a latticed network of qubits with nearest-neighbor interactions. Superconducting qubits have continued to steadily improve in coherence, gate, and readout fidelities, to become a leading candidate for implementation into larger quantum networks. Here we describe characterization experiments and calibration of a system of four superconducting qubits arranged in a planar lattice, amenable to the surface code. Insights into the particular qubit design and comparison between simulated parameters and experimentally determined parameters are given. Single- and two-qubit gate tune-up procedures are described and results for simultaneously benchmarking pairs of two-qubit gates are given. All controls are eventually used for an arbitrary error detection protocol described in separate work [\emph{Corcoles et al., Nature Communications, 6, 2015}].
\end{abstract}

%>>>> Include a list of keywords after the abstract 


%%%%%%%%%%%%%%%%%%%%%%%%%%%%%%%%%%%%%%%%%%%%%%%%%%%%%%%%%%%%%
\section{INTRODUCTION}
\label{sec:intro}  % \label{} allows reference to this section

To deal with the inherently fragile nature of physical qubit systems, quantum error correction~\cite{Shor1995} protocols are required to make fault-tolerant quantum computing a reality. Superconducting qubits have begun to enter a phase of study where the challenge is to continue to build larger inter-connected networks for realization of quantum error correction protocols. The surface code~\cite{Bravyi1998,Raussendorf2007,fowler_surface_2012} is one particularly attractive fault-tolerant quantum error correction architecture, requiring only nearest-neighbor interactions, and physically realized through a simple square lattice arrangement. Furthermore, threshold error levels for the surface code are low at $\sim$1\%, which are reasonable targets for superconducting qubit systems which have continued to show steady progress in coherence times~\cite{Paik2011,chang_improved_2013,Devoret2013}. 

Pairing the ideas of the surface code error correction protocol with superconducting qubits has already resulted in multiple important demonstrations. Ancilla-based bit-flip parity detection in networks of three superconducting qubits have been shown and characterized \cite{Saira2014,Chow2014}. Dealing with classical bit-flip parity errors was taken a step further through the implementation of a 5 and 9-qubit repetition code \cite{Kelly2015} in a linear network of superconducting transmon qubits, successfully showing increased preservation of an encoded bit-flip state with larger code distance. A true quantum code was implemented in a 4-qubit two-by-two lattice arrangement of transmons, with the ability to detect an arbitrary single-qubit quantum error to a codeword of two code qubits, via measurement of two ancilla qubits, one for bit-flip errors, and one for phase-flip errors~\cite{Corcoles2015}. In this Proceeding, we expand on that particular work by going into further detail on device design and simultaneous gate characterization in the 4-qubit lattice experiment from Ref.~\citenum{Corcoles2015}.
%%%%%%%%%%%%%%%%%%%%%%%%%%%%%%%%%%%%%%%%%%%%%%%%%%%%%%%%%%%%%
\section{Device Design and Parameters} 

\begin{figure}
	\centering
	\includegraphics[width=5in]{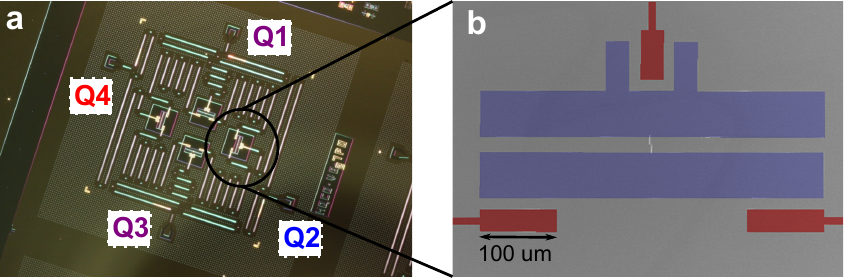}
	\caption{\label{fig:Fig1} \textbf{a}. Optical micrograph of two-by-two lattice device. Four qubits are labeled, with Q1 and Q3 serving as code qubits and Q2 and Q4 serving as either bit-flip or phase-flip error syndrome qubits. This is the same device studied in Ref.~\citenum{Corcoles2015}. The device contains four quantum bus resonators which couple nearest neighbor qubits, as well as four additional readout resonators, one per qubit, for independent determination of the qubit state and for independent addressability. \textbf{b}. False-colored zoom-in optical micrograph of the transmon qubit used in the four-qubit lattice device. The transmon islands are colored in blue, connected by the Josephson junction in the middle. The red-colored arm on the top of the transmon structure defines the coupling capacitance to the readout resonator, while the two red-colored arms on the left and right bottom of the transmon structure define the coupling capacitance to two bus resonators for coupling to neighboring transmons. A scale bar is included for dimensional reference.}
\end{figure}

We construct a four-qubit lattice using four superconducting transmon qubits~\cite{koch_charge-insensitive_2007}, connected in a ring arrangement with four coplanar waveguide quantum bus resonators~\cite{majer_coupling_2007}, and each transmon qubit has an independent readout resonator. Every quantum bus connects two qubits, and every qubit is connected to two buses. The four-qubit device is shown in Fig.~\ref{fig:Fig1}a with labels for each of the four qubits. The red (Q4) and blue (Q2) qubits are used for syndrome error detection, whilst the two purple qubits (Q1 and Q3) are code qubits. The four-qubit device is a small modification to the three-qubit half-plaquette device of Ref.~\citenum{Chow2014}. The major differences are the addition of one transmon, an additional readout resonator for that transmon, and two additional bus resonators for completing the lattice connectivity. We next describe the design of this transmon, along with reasoning for targeting particular circuit parameters and qubit properties. That is followed by a section describing experimentally obtained device parameters for the four-qubit device that we have studied, and we include a discussion about the comparison with the targeted simulation values.

%%-----------------------------------------------------------
\subsection{Transmon design and parameter selection} 
\label{sec:transmondesign}

An optical micrograph of the superconducting transmon qubit design used in both Refs.~\citenum{Chow2014} and \citenum{Corcoles2015} is shown in Fig.~\ref{fig:Fig1}b. The particular parameter regime which we aim for is strongly influenced by the need for high quality controls and readout of the qubits. 

One key aspect of constructing our devices is to preserve the ability to perform high-fidelity two-qubit gates. Although there are a host of different implementations of two-qubit gates for superconducting qubits, the one which we focus on is the cross-resonance (CR) gate. The CR gate has the advantages that it can be performed with microwave control, and does not require dynamic tunability of the qubit frequencies. Details about the CR gate and interaction have been published previously~\cite{Paraoanu2006,Rigetti2010,Chow2011}, but functionally, there are two labeled qubits, a control qubit and a target qubit. A microwave drive onto the control qubit at the target transition frequency, will result in a control qubit state dependent rotation of the target qubit. Via proper calibration of this rotation, it is possible to realize either a controlled-NOT operation, or a $\pi/2$ rotation around the $ZX$ two-qubit axis ($ZX_{\pi/2}$, where the notation follows with the control-qubit labeled first), which serves as the primitive for constructing gates from the two-qubit Clifford group~\cite{Corcoles2011}.

Detailed experimental and theoretical study of the CR gate~\cite{ware_inprep_2015} with transmon qubits have shown that there is an optimal detuning window between the control and target transmon qubit frequencies which results in stronger interactions and hence faster gates. This detuning window corresponds to placing the target transmon transition frequency right between the two frequencies corresponding to the lowest two energy level differences of the control transmon. The devices we work with are fixed-frequency transmons (no SQUID loop for magnetic flux tunability), so as to avoid decoherence due to residual flux noise~\cite{harlingen_decoherence_2004}. With fixed-frequency transmon devices, hitting this window comes down to having accurate control over the transmon capacitance values and the critical currents of the Josephson junctions. To that end, through multiple iterations of device designs and experimental extraction of transmon parameters, it has been observed that the primary cause of frequency fluctuations in nominally identical designs is critical current fluctuations on the order of $\sim 10\%$, and to a lesser degree capacitance variation. As such, in order to design larger networks with many transmons that preserve the ability to perform reliable and fast CR gates between neighboring transmons, it is necessary to make this anharmonicity window large enough that even with the variation in transmon frequency from fabrication, that we will fall within the window. Future work will be aimed at minimizing these critical current fluctuations in order to avoid frequency collisions in even larger arrays of qubits.

For our four-qubit devices, we settle on a transmon capacitance that results in $\sim 340$ MHz of anharmonicity. With lowest energy level frequencies of $\sim 5$ GHz, this means with typical critical current fluctuations that we see on order of $\sim 400-500$ MHz of frequency spread, and that neighboring transmon devices will have a reasonable chance at hitting the CR gate window given their relative detuning. With this level of anharmonicity, it is also important to remain strictly around $\sim 5$ GHz, otherwise, any lower will result in the device not behaving in the transmon regime~\cite{koch_charge-insensitive_2007}.

By narrowing down the transmon transition frequencies ($\sim$5 GHz) and transmon anharmonicities ($\sim 340$MHz), we are guided towards the selection of the readout resonator frequencies and quantum bus frequencies. In order to have high-fidelity readout, it is important to get in the regime where standard quantum-limited amplifiers can function. To pair specifically with Josephson parametric amplifiers~\cite{johnson_heralded_2012} and superconducting low-inductance undulatory galvanometer amplifiers~\cite{Hover2012}, we choose readout frequencies around 6.4-6.8 GHz, which fall within the tunable bandwidth for gain with such amplifiers. We then carefully select an appropriate coupling strength $g_{\text{R}}$ and readout resonator coupling capacitance $C_{\text{R}}$ so as to satisfy  optimal readout fidelity conditions (as described in Ref.~\citenum{Magesan2014}) and yet not be overly limited via Purcell-loss through the resonator~\cite{houck_purcell_2008}. Future work will entail integrating Purcell filters to permit faster readouts with stronger couplings, while avoiding spontaneous decay via the coupled resonator~\cite{Reed2010, Jeffrey2014,Bronn2015}. Similar considerations are taken into account for the bus resonators, with the added caveat that each transmon device must couple to two independent bus resonators. To avoid frequency collision between the readout resonators and the bus resonators, we choose 7.6 and 8.1 GHz for neighboring buses. The coupling strength to each transmon is chosen to be relatively large, $g\sim 80-90$ MHz, so as to achieve strong CR interactions. With the transmons at 5 GHz, loss due to spontaneous emission into the bus cavities is also suppressed. Table \ref{table:1} summarizes the targeted design values for various parameters for our devices. Capacitance simulations are done using ANSYS Q3D and we use network theory for a reduction to a minimal model.

\begin{table}
	\centering
		\begin{tabular}{|c|c |c c c c|}
			\hline
			Quantity & Targeted & Q1  & Q2 & Q3 & Q4\\ 
			\hline
			qubit transition frequency (GHz) & 5.3 & 5.303 & 5.101 & 5.291 & 5.415 \\
			anharmonicity (MHz) & -339.9 & $-340\pm3$ & $-340\pm3$ & $-341\pm3$ & $-340\pm3$ \\
			critical current (nA) 	& 27  	& 26.8 (27.2) 	& 25.1 (25.4) & 26.7 (27) & 27.8 (28.2)\\%_2;09
			qubit capacitance (fF) 	& 62 + $C_{\text{J}}$	& 65.5 (66.5)  	& 65.9 (66.9) & 65.3 (66.3)&65.3 (66.3)\\ %_1;09
			$E_{\text{J}}/E_{\text{C}}$	& 45.7	& 45.0		& 42.4& 44.7& 46.5\\%_1;09
			charge dispersion (kHz)	& 24.9 & 28.3	&	45.3 	& 30.0	& 21.6\\%_2;09
			$T_{\phi}$ from charge (ms) & 41 & 35.8 & 22.4 & 33.8 & 47.0 \\
			readout resonator (GHz) & 6.5/6.7& 6.494 & 6.695 & 6.491 & 6.693\\ 
			readout $Q$ factor & 15000 & 10560 & 15200 & 22600 & 5530 \\
			dispersive shift $\chi$ (MHz) & -1.6 & -1.5 & -1.0 & -1.25 &  -1.4 \\
			$g_{\text{R}}$ coupling to readout (MHz) & 94 & 89 & 94 & 82 & 92\\
			coupling capacitance $C_{\text{R}}$ (fF) & 5.5 & 5.4 & 5.7 & 5.0 & 5.3 \\
			Purcell limited $T_1$ ($\mu$s) & 69 & 70 & 182 & 182  & 40 \\
			\hline	 
		\end{tabular}
	\caption{\label{table:1} Summary and comparison of parameters. We present designed values of a number of transmon parameters and transmon to readout resonator coupling parameters. Other columns include experimentally extracted and inferred parameters for comparison to the designed values. $C_{\text{J}}$ refers to the capacitance of the Josephson junction, which is typically $\sim$2-3 fF. A key result of the comparison of simulation and experiment is that the qubit capacitance comes out very close to what is designed. Future iterations of design with experimental extraction of parameters will aim to further tighten the bounds between simulation and experiment. Part of this comes down to more accurately determining certain parameters from the experiment, for example, $\chi$, to avoid the deviations observed in quantities like $C_{\text{R}}$. We believe variations in the $Q$ factor observed in the experiment are due to differences in the impedances on the external lines which drive the cavity outside of the sample device.}
\end{table}

\begin{figure}
	\centering
	\includegraphics[width=5in]{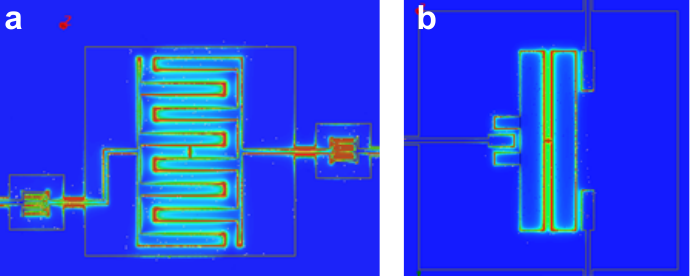}
	\caption{\label{fig:Fig2} Ansoft HFSS electric field density simulations of two different transmon geometries. A 30 $\mu$m finger by 30 $\mu$m gap interdigitated capacitor is shown in \textbf{a} and the parallel pad layout as employed in the four-qubit lattice is shown in \textbf{b}. From these simulations it is possible to extract field participation of specific targeted interfaces between the two geometries. Analysis of these two geometries reveals a significant reduction in the participations of the substrate-air, metal-air, and substrate-metal interfaces in $\textbf{b}$ compared to $\textbf{a}$. Previous work~\cite{Corcoles2011,Chow2012PRL} employing the design from \textbf{a} shows considerably lower transmon quality factors when compared to experiments employing the parallel pad design~\cite{Chow2014,Corcoles2015}. Future study of transmon designs will explore targeting specific interfaces in the design to determine which are the most critical for further reducing surface loss and improving coherence times. Note the size of the device in \textbf{b} is the same as in Fig.~\ref{fig:Fig1}b. }
\end{figure}

As seen in Fig.~\ref{fig:Fig1}b, the transmon design has a very simplified parallel pad layout. This design departs from previous transmon schemes in which the capacitance is defined via interdigitated finger electrodes~\cite{Corcoles2011,chang_improved_2013}. The avoidance of finger capacitors for the transmon capacitance and for any coupling capacitance to bus or readout resonators is a conscious decision made to abate susceptibility and participation of lossy interfaces~\cite{Wenner2011a,chang_improved_2013,Barends2013}. As an example, Fig.~\ref{fig:Fig2} shows two transmon designs which have been studied in our group. The first, Fig.~\ref{fig:Fig2}a consists of a main transmon capacitor made of 30 $\mu$m fingers with 30 $\mu$m gaps. This device is similar to ones studied in single-qubit modalities as in Ref.~\citenum{chang_improved_2013}. In those devices, it was observed phenomenologically that going to a larger finger and gap width for the interdigitated capacitance improved the quality factor of the underlying qubits, consistent with reducing the participation of substrate-air and substrate-metal interfaces in surface loss simulations. In addition, the old designs had a considerable part of the qubit capacitance in the couplers to the resonators.  We subsequently moved to the current design as in Fig.~\ref{fig:Fig2}b, which provides the capacitance via a parallel pad geometry and reduces the participation of the qubit capacitance in the couplers to the resonators.  

The colored density plots in Fig.~\ref{fig:Fig2} are ANSYS HFSS simulated electric-field densities. Simulations of the electric-field participation  in the substrate-air, metal-air, and substrate-metal interfaces are determined. To compare the two different geometries of Fig.~\ref{fig:Fig2}, we find that the parallel pad configuration results in a reduction of the substrate-air interface participation by $\sim 35\%$, metal-air interface participation by $\sim 44\%$, and substrate-metal interface participation by $\sim37\%$.  We attribute increased transmon quality factors to these changes to the design, but at this moment, we cannot yet pinpoint which of these are the most dominant factor for improvement. 

\subsection{Four-qubit device parameters and comparison to simulation}
From our experiment with the four-qubit device, we find transition frequencies for Q1, Q2, Q3, and Q4 at 5.303, 5.101, 5.291, and 5.415 GHz, respectively  along with corresponding measured anharmonicities all around $-340\pm3$ MHz. From measurement of the interaction with the readout cavities, dispersive shift $\chi/2\pi$ is found to be -1.5,-1.0,-1.25, and -1.4 MHz . Using these measured values, we complete the values of Table~\ref{table:1} with a comparison to the targeted values from simulation. We can see that for this device, in which the four qubits are fabricated with the same targeted design values, there is still a considerable spread in the qubit frequencies. Nonetheless, all qubits fall within the anharmonicity window which make it usable for nearest-neighbor two-qubit CR gates. 

The readout resonators are designed at 6.5 and 6.7 GHz, and the experiment shows excellent agreement for these designs. There is some observed variation in the measured quality factors of the four resonators, which results in some differences in the expected Purcell decay channels for the four different qubits~\cite{houck_purcell_2008}. Future designs of these lattice devices will employ Purcell filtering~\cite{Reed2010} so as to allow for fast readout and still preserve qubit lifetimes. Nonetheless, good agreement between simulations and our actual experiments are important for further optimizing geometries to be incorporated into larger lattices of qubits for the surface code. 

%%%%%%%%%%%%%%%%%%%%%%%%%%%%%%%%%%%%%%%%%%%%%%%%%%%%%%%%%%%%%
\section{Calibrating Controls}
Through standard sliding $\pi$-pulse and spin echo sequences, we characterize the coherence times of the four-qubit device. We find energy relaxation times $T_1$= $33.2\pm2.4$, $36.0\pm5.7$, $30.8\pm2.7$, $28.6\pm3.6$ $\mu$s and coherence times $T_2^{\rm{echo}}$ = $16.9\pm1.4$, $16.0\pm3.5$,$18.2\pm1.9$,$22.2\pm3.3$ $\mu$s for Q1, Q2, Q3,and Q4 respectively. These experiments are performed repeatedly over the course of two days, to gather statistics and average values for the coherence times. Subsequently, we calibrate single-qubit gates for each of the four qubits as well as two-qubit CR gates between nearest neighbor qubits in the lattice.

All microwave pulses are applied using single-sideband (SSB) modulation of a carrier microwave tone using an arbitrary waveform generator. Using SSB ensures a well-defined phase relationship between the two quadratures for performing single-qubit rotations either around the $x$-axis or $y$-axis and avoids carrier leakage on resonance with the qubit transition. The single-qubit gates that are calibrated in our system include $\pi/2$ and $\pi$ rotations around the $x$- and $y$-axes of each qubit. First, the $\pi/2$ amplitudes are tuned up by measuring the cavity response after $\{X_{\pi/2} ( X_{\pi/2} X_{\pi/2})^{N}\}$ for $N$ up to 7. Then, we bootstrap off the tuned up $\pi/2$ pulse to find the appropriate $\pi$ amplitude via the pulse train $\{X_{\pi/2}(X_{\pi})^N\}$, again for $N$ up to 7 pulses. Finally, the derivative-pulse shaping (DRAG) parameter~\cite{Motzoi:2009fx} is tuned via the sequence $\{X_{\pi/2}\left(X_{\pi/2}X_{-\pi/2}\right)^N\}$. Further study into the accurate calibration of these single-qubit gates is the subject of separate work~\cite{Sheldon2015}. 

The two-qubit CR gates are calibrated in close analogy to single qubit amplitude calibrations. Following the refocusing sequence as described in Ref.~\citenum{Corcoles2011}, it is possible to define a $ZX_{\pi/2}$ two-qubit gate. An odd number $2N-1$ of $ZX_{\pi/2}$ pulses are applied and the amplitude is adjusted so that for each $N$ the expected signal is halfway between 0 and 1 (in our experiment $N$=5). Any amplitude miscalibrations lead to departures from this expected signal and are amplified for increasing $N$.

In addition to amplitude we must also calibrate the phase of the $ZX_{\pi/2}$ pulse of the CR gate to the phase of the target qubit. The phase of the complete microwave pulse realizing the CR gate is tuned to match that of the microwave pulse which realizes the single qubit rotations on the target. This is done by applying the pulse sequence~\cite{Chow2014} $IY_{\pi/2}(ZU_{\pi}IX)^N IX_{\pi/2}$. The $U$ denotes the rotation axis defined by the CR pulse and the goal is to calibrate to match an $x$ rotation on the target qubit. In the case of an $x$-rotation we expect the signal to be halfway between 0 and 1 for each $N$ and miscalibrations of the phase lead to deviations that are amplified with increasing $N$. 

\section{Simultaneous Gate Characterization and Validation}
\begin{figure}
	\centering
	\includegraphics[width=5in]{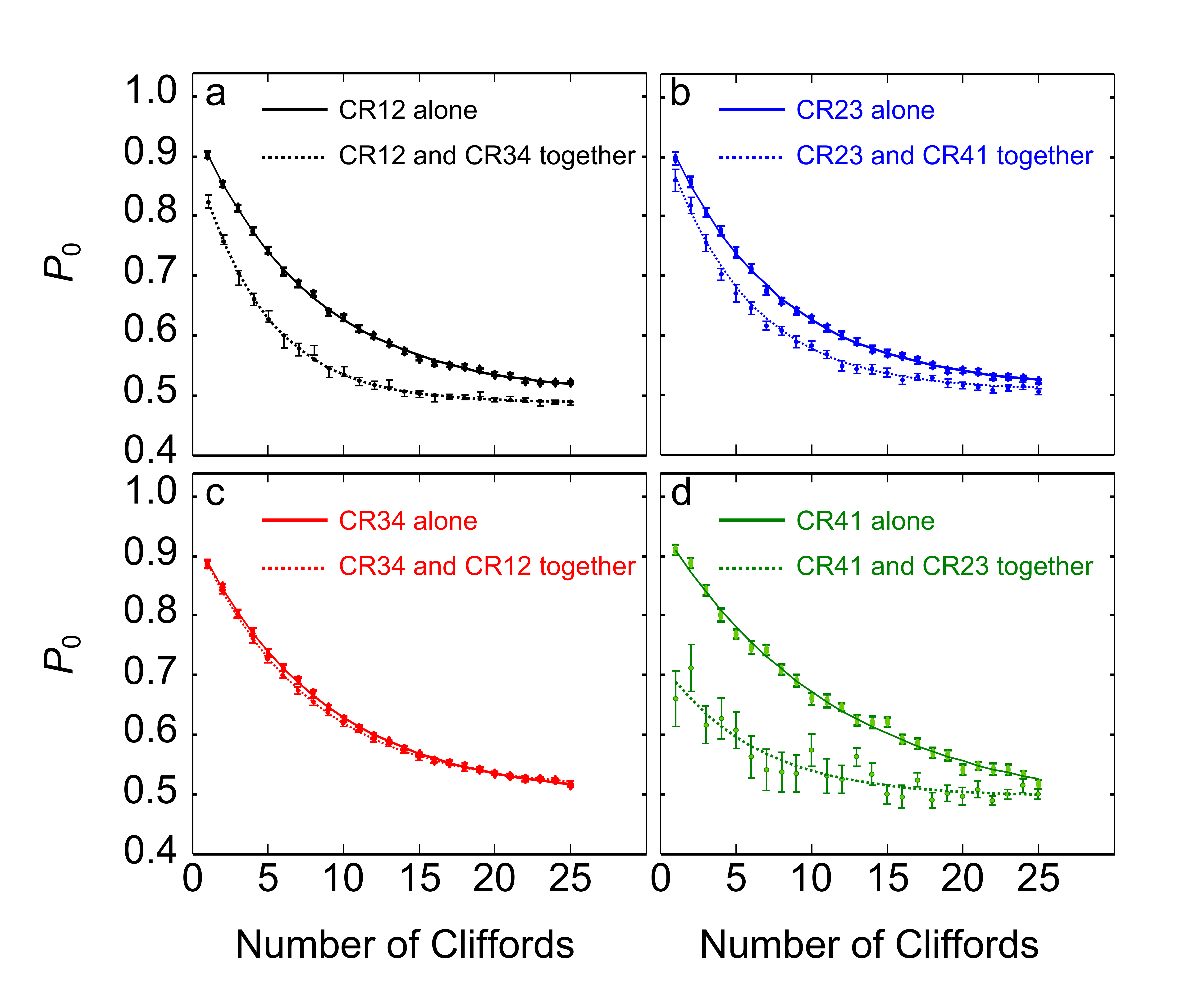}
	\caption{\label{fig:Fig3} Standard and simultaneous two-qubit Clifford randomized benchmarking, indicated as average sequence fidelity versus number of Clifford gates. All four nearest-neighbor two-qubit CR gates are calibrated and then characterized to obtain an average gate fidelity via Clifford randomized benchmarking over 30 different randomization sequences. The technique follows that outlined in Ref.~\citenum{Corcoles2011}. The gate fidelities extracted are $0.9396\pm0.0006$, $0.9369\pm 0.0007$, $0.9431\pm0.0015$, $0.9647\pm0.0015$ for CR12 (\textbf{a}), CR23 (\textbf{b}), CR34 (\textbf{c}), and CR41 (\textbf{d}), respectively. Subsequently, simultaneous two-qubit Clifford benchmarking is also run, with CR12 and CR34 at the same time, and CR23 and CR41 at the same time. The average fidelities in these simultaneous cases is a fair bit lower for three of the four gates, with $0.9017$ and $0.9367$ for CR12 and CR34, and 0.9149 and 0.9241 for CR23 and CR41. The origin of the fidelity reduction is the topic of on-going study. It will be necessary to determine how much of this error is due to classical crosstalk and how much is from actual undesired Hamiltonian effects~\cite{ware_inprep_2015}. }
\end{figure}

Having calibrated the single- and two-qubit gates for all qubits and all nearest neighbor CR operations, we apply the techniques of Clifford randomized benchmarking~\cite{Magesan2011} (RB) to characterize the quality of the calibrated gates through an average gate fidelity. In this technique, random sequences of gates chosen from the Clifford group are constructed, paired with a final Clifford gate which returns the qubit to the ground state. Both single and two-qubit Clifford benchmarking are performed. Average sequence fidelity of this ground state versus an increasing number of Clifford gates is then plotted, and performed over many different randomized sequences of Cliffords. For single-qubit gates, we calibrate pulses of length 53.3 ns, and find average gate fidelities of $0.9986$, $0.9988$, $0.9988$, $0.9989\pm0.0003$ for Q1, Q2, Q3, and Q4 respectively. We also characterize the nearest-neighbor two-qubit CR gates via RB~\cite{Corcoles2011}. The averaged two-qubit RB sequence fidelity versus the number of Cliffords over 30 different randomizations is shown in Fig.~\ref{fig:Fig3} as the solid curves. From these curves, we extract gate fidelities of $0.9396\pm0.0006$, $0.9369\pm 0.0007$, $0.9431\pm0.0015$, $0.9647\pm0.0015$ for CR12, CR23, CR34, and CR41, respectively. Here the notation $\text{CR}ij$ refers to a CR gate between Q$i$ and Q$j$, with $i$ referencing the control qubit and $j$ referencing the target qubit. 

Another important part of characterizing the gates for this four-qubit device is to look at crosstalk effects. To tackle this problem, we employ the technique of simultaneous RB~\cite{gambetta_characterization_2012}. By simultaneously applying randomized Clifford sequences to multiple qubits it is possible to get a bound on the amount of crosstalk errors in the system. A full table showing the effect of simultaneous RB, with all combinations of simultaneous application of RB experiments for the four qubits, is given in Ref.~\citenum{Corcoles2015}. Comparing the individual and simultaneous RB experiments,it is observed that the addressability error is 0.001 or lower in all cases for our single-qubit gates.

It is also possible to check simultaneous RB of the two-qubit gates. As the two-qubit gates involve two qubits at a time, it is only possible to run two sets of simultaneous gates: CR12 with CR34 at the same time, and CR23 with CR41 at the same time. The average sequence fidelity versus Clifford number in these simultaneous cases are also shown in Fig.~\ref{fig:Fig3} as the dashed lines.  When CR12 and CR34 are simultaneously benchmarked, we now obtain fidelities of $0.9017$ and $0.9367$ for CR12 and CR34, respectively. CR12 has significantly worsened in this experiment, although CR34 is not very affected. Similarly, for CR23 and CR41, we now obtain fidelities of 0.9149 and 0.9241, respectively, which are both significantly lower than when the same gates are characterized independently. 

It is clear that in three of the four two-qubit gate cases, the simultaneous application of the gates alter and degrade their performance. The exact nature of this crosstalk is the subject of further study and understanding it will help us control and construct larger lattices. Nonetheless, even with this level of addressability error, the quantum error detection experiments as described in Ref.~\citenum{Corcoles2015} can still be run because many of the gates in that protocol are staggered and not run at the same time. Future work will address how to make these controls more robust to simultaneous operation, which may be necessary in more complicated algorithms.

\section{Conclusions}
In this Proceeding, we presented additional detail on the design rules for transmon qubits in the four-qubit lattice device used for arbitrary error detection in Ref.~\citenum{Corcoles2015}. We showed the comparison between designed parameters and experimentally extracted parameters. We also presented results of simultaneous application of control gates in the device. Looking ahead, the further exploration of addressability errors arising from two-qubit gates in a lattice of qubits will be critical to understanding how to control larger lattices of qubits for the surface code quantum error correction protocol. 

%%%%%%%%%%%%%%%%%%%%%%%%%%%%%%%%%%%%%%%%%%%%%%%%%%%%%%%%%%%%%
\acknowledgments     %>>>> equivalent to \section*{ACKNOWLEDGMENTS}       
 
	We thank M.~B.~Rothwell and G.~A.~Keefe for fabricating devices. We thank J.~R.~Rozen, J.~Rohrs and K. Fung for experimental contributions. We acknowledge support from IARPA under contract W911NF-10-1-0324. All statements of fact, opinion or conclusions contained herein are those of the authors and should not be construed as representing the official views or policies of the U.S. Government.

%%%%%%%%%%%%%%%%%%%%%%%%%%%%%%%%%%%%%%%%%%%%%%%%%%%%%%%%%%%%%
%%%%% References %%%%%

\end{document}